\documentclass[reprint, NumberedRefs]{JASA}
\usepackage{comment}
\usepackage{soul}
\usepackage{siunitx}
\usepackage{tikz}
\usepackage{pgfplots}
\pgfplotsset{compat=1.12}
\usetikzlibrary{plotmarks}
\definecolor{blue}{rgb}{0.121, 0.462 , 0.705}
\definecolor{red}{rgb}{0.839, 0.152, 0.156}
\definecolor{hard_red}{rgb}{1, 0, 0}
\definecolor{orange}{rgb}{1, 0.501, 0.058}
\definecolor{brown}{rgb}{0.549, 0.337, 0.294}
\newcommand{\bluelegend}{\raisebox{2pt}{\tikz{\draw[blue, solid, line width = 1pt](0,0) -- (3mm,0);}}}
\newcommand{\redlegend}{\raisebox{2pt}{\tikz{\draw[red, solid, line width = 1pt](0,0) -- (3mm,0);}}}
\newcommand{\orangelegend}{\raisebox{2pt}{\tikz{\draw[orange, solid, line width = 1pt] (0,0) -- (3mm,0);}}}
\newcommand{\brownlegend}{\raisebox{2pt}{\tikz{\draw[brown, solid, line width = 1pt](0,0) -- (3mm,0);}}}

\begin{document}

\title[]{}

\title[JASA/Sample JASA Article]{Sound field reconstruction in rooms: inpainting meets super-resolution}

\author{Francesc Llu\'is}
\email{lluis-salvado@mdw.ac.at}
\affiliation{Department of Music Acoustics, University of Music and Performing Arts Vienna, Austria}
\author{Pablo Mart\'inez-Nuevo}
\author{Martin Bo M\o ller}
\author{Sven Ewan Shepstone}
\affiliation{R\&D Acoustics,  Bang \& Olufsen, Struer, 7600, Denmark}

\preprint{Author, JASA}		

\date{\today}

\begin{abstract}
In this paper, a deep-learning-based method for sound field reconstruction is proposed. It is shown the possibility to reconstruct the magnitude of the sound pressure in the frequency band 30-300 Hz for an entire room by using a very low number of irregularly distributed microphones arbitrarily arranged. Moreover, the approach is agnostic to the location of the measurements in the Euclidean space. In particular, the presented approach uses a limited number of arbitrary discrete measurements of the magnitude of the sound field pressure in order to extrapolate this field to a higher-resolution grid of discrete points in space with a low computational complexity. The method is based on a U-net-like neural network with partial convolutions trained solely on simulated data, which itself is constructed from numerical simulations of Green's function across thousands of common rectangular rooms. Although extensible to three dimensions and different room shapes, the method focuses on reconstructing a two-dimensional plane of a rectangular room from measurements of the three-dimensional sound field. Experiments using simulated data together with an experimental validation in a real listening room are shown. The results suggest a performance which may exceed conventional reconstruction techniques for a low number of microphones and computational requirements.
\end{abstract}


\maketitle

\maketitle

\section{\label{sec:1} Introduction}
The functions describing sound propagation, such as sound pressure or particle velocity, operate scalar and vector values respectively which vary across the temporal and spatial dimensions. There are many applications where knowledge of the spatial variation of the sound field is of paramount interest, for example, sound field navigation for virtual reality environments \citep{Tylka:2015aa,Tylka:2016aa}, accurate spatial sound field reproduction over predefined regions of space \citep{Berkhout:1993aa,Druyvesteyn:1997aa,Ward:2001aa}, or sound field control in reverberant environments \citep{Betlehem:2005aa,Radlovic:2000aa}.

The different reconstruction scenarios are determined by the type of information gathered from the sound field. Depending on the type of acquisition, several techniques are used, ranging for example, from acoustic holography \citep{Williams:1999ab}, acousto-optic methods \citep{Torras-Rosell:2012aa,fernandez2013holographic}, or traditional discrete sets of spatial samples \citep{Ajdler:2006aa}. The latter is particularly convenient in practice since it requires simple microphones.

In the case of sound field reconstruction in rooms, there exist several methods in the literature. In particular, model-based approaches based on samples of the sound pressure at a discrete set of locations tend to dominate the area. Results using classical sampling \citep{Ajdler:2006aa}, i.e. based on bandwidth analysis, build upon the image source method to characterize the sound field in a room in order to derive bounds on the aliasing error for a given sampling density. This leads to an impractically high density of microphones for an acceptable reconstruction error. Another approach to simplify the model and the number of measurements is based on parameterizing the room impulse response as a pole-zero system \citep{Haneda:1999aa}.

Compressive sensing approaches have been effective in reducing the number of measurements compared to these previous methods. They inherently require an underlying assumption of the sparsity of the chosen room acoustics model. Utilizing modal theory, it is possible to consider a plane wave approximation of the sound field \cite{Moiola:2011aa} in a room in order to describe it spatially as a sparse linear combination of damped complex exponentials \cite{Mignot:2013ab,Antonello:2017aa,Verburg:2018aa}. Dictionaries tend to be large, performance degrades at high frequencies, and the interpolated location should be, in general, in the far field with respect to the source. Under the image source method, estimation of the early part of the room impulse response is also possible assuming a few dominant image sources \cite{Mignot:2013aa}. These techniques are in general sensitive to the choice of sampling scheme used in order to guarantee meaningful solutions and well-conditioned problems. Empirical methods for the latter are commonly adopted leading to some restrictions in the arrangement of microphones. Exploiting information about the modal frequencies may allow a more general microphone arrangement \cite{grande2019sound} at the expense of sensitivity to source location, modal density, and accurate modal frequencies estimation. Additionally, finding solutions to these sparse inverse problems is typically computationally demanding \cite{Kim:2007aa}.

In this paper, we adopt a data-driven approach to the problem of sound field sampling and reconstruction, which, for the present application, appears to be unexplored. For clarity of exposition, we focus on a two-dimensional horizontal plane of three-dimensional rectangular rooms. We consider a very low number of irregularly and arbitrarily distributed measurements to recover the magnitude of the sound pressure in a room across the spatial dimension for the frequency range 30-300 Hz. In contrast to previous methods, our approach is location agnostic in the sense that it does not require knowledge of the microphone positions or the interpolation points in the Euclidean space. These characteristics can contribute to designing more practical sampling and reconstruction procedures. The goal of the paper is then threefold: use a very low number of microphones, accommodate irregular and location agnostic microphone distributions, and carry out inference that is computationally efficient.

We first view the sound field as a two-dimensional discrete signal. The acquisition step can be interpreted as producing a low-resolution signal with missing samples. Then, the recovery step consists of filling the missing data of a high-resolution two-dimensional signal. We show how this process can be viewed as jointly performing inpainting \cite{bertalmio2000image,Liu:2018aa} and super-resolution \cite{freeman2002example,zhang2018residual}, both well-known techniques in image processing with a good performance using deep learning methods. In particular, we use a U-net neural network \cite{Ronneberger:2015aa} with partial convolutions \cite{Liu:2018aa} trained on simulated data that simultaneously performs inpainting and super-resolution. Under this framework, we show how it is possible to recover a high-resolution field from a very low number of irregular and location-agnostic measurements with low computational complexity in the inference process.

The paper is organized as follows: Section \ref{sec:ProblemDescription} establishes the conceptual framework under which the reconstruction problem is addressed, i.e. as a learning algorithm drawing upon inpainting and super-resolution techniques. The details about the neural network architecture and the training procedure used for recovery are explained in Section \ref{sec:Approach}. Section \ref{sec:Results} presents results concerning the reconstruction accuracy of the proposed algorithm both in simulated and experimental settings, i.e. in real rooms.

\section{\label{sec:ProblemDescription} Problem Description}
We frame the problem of sound field reconstruction within a data-driven approach, i.e. we aim at developing a recovery algorithm that directly and progressively learns from raw sound field data. The machine learning methods that have been particularly successful in this regard fall under deep learning systems. These have significantly outperformed model-based approaches in tasks such as, but not limited to, image classification, analysis, and restoration \cite{he2016deep, chollet2017xception}; or speech recognition and synthesis\cite{nassif2019speech, wang2017tacotron}.

The novelty of the present approach lies in the observation that the magnitude of the sound pressure in a room can be interpreted as a two-dimensional discrete function defined on a rectangular grid of points in space, i.e. in the same way a raster image is represented by a rectangular grid of pixels. This allows us to exploit the effectiveness of deep learning techniques in image processing. Although the principles governing the proposed algorithm can, in principle, be extended to three-dimensional regions, we focus on reconstructing the three-dimensional field in a two-dimensional plane for the sake of simplicity. We further assume that the enclosures of interest consist of rectangular rooms corresponding to domestic standards \cite{international2015recommendation}. Note that the method described here could also be extended to different room shapes.

In particular, the function that we sample and reconstruct is a discrete version of the magnitude of the Fourier transform of the sound field in a given frequency band. We show in the following how reconstructing this function is connected to the well-known concepts of image inpainting and super-resolution. Let us first denote the spatio-temporal sound field in a three-dimensional rectangular room as $p(\mathbf{r},t)$ where $\mathcal{R}=(0,l_x)\times(0,l_y)\times(0,l_z)$ for some $l_x,l_y,l_z>0$ and $\mathbf{r}\in\mathcal{R}$. The magnitude of its Fourier transform is given by
\begin{equation}
s(\mathbf{r},\omega):=\Big|\int_\mathbb{R}p(\mathbf{r},t)e^{-j\omega t}\mathrm{d}t\Big|
\end{equation}
 for $\omega\in\mathbb{R}$ and $\mathbf{r}\in\mathcal{R}$. 

Initially, given a room, we can define the following rectangular grid as a set on an arbitrary two-dimensional plane, i.e.
\begin{equation}
\mathcal{D}_o:=\Big\{\Big(i\frac{l_x}{I-1},j\frac{l_y}{J-1},z_o\Big)\Big\}_{i,j}
\end{equation}
for $z_o\in(0,l_z)$, $i=0,\ldots,I-1$, $j=0,\ldots,J-1$, and some integers $I,J\geq2$. Then, the available spatial sample points, denoted as $\mathcal{S}_o$, consist of a subset of $\mathcal{D}_o$. It is important to observe that there is no constraint whatsoever with regard to the pattern that $\mathcal{S}_o$ has to form within $\mathcal{D}_o$. This allows us to have, for example, irregularly distributed spatial sample points within the room. For a given excitation frequency, the available samples can then be expressed as follows
\begin{equation}
\{s(\mathbf{r},\omega)\}_{\mathbf{r}\in\mathcal{S}_o\subseteq\mathcal{D}_o}.
\end{equation}
Note that the problem of interpolating $s(\mathbf{r},\omega)$ to the entire domain $\mathcal{D}_o$ from known values in $\mathcal{S}_o$ can be viewed as image inpainting, i.e. filling in the missing holes of a raster image. This is motivated by the irregular nature of the sampling pattern.

\begin{figure}[!t]
	\centering
	\includegraphics[scale=0.7]{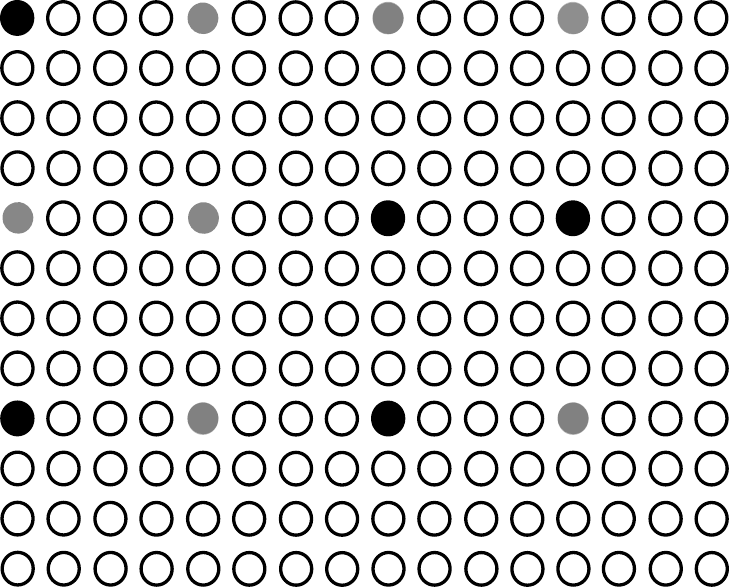}
	\caption{Illustration of the spatial points considered for reconstruction of the function $s(\mathbf{r},\omega)$ for a given frequency. The set $\mathcal{D}_o$ consists of the solid black and gray circles where the former, for example, can be interpreted as $\mathcal{S}_o$. The set $\mathcal{D}_o^{L,P}$ is then given by all the points depicted where inpainting and super-resolution is jointly performed from knowledge of the function in $\mathcal{S}_o$. Note that here $L=P=4$. (Color online.)}
	\label{fig:sr_inpainting}
\end{figure}

However, we are interested in reconstruction on an even finer rectangular grid in order to capture the small-scale spatial variations of the sound field. In order to do so, we eventually interpolate the sound field to a grid of points corresponding to an upsampled version of the set $\mathcal{D}_o$, i.e. 
\begin{equation}
\mathcal{D}_o^{L,P}:=\Big\{\Big(i\frac{l_x}{(I-1)L},j\frac{l_y}{(J-1)P},z_o\Big)\Big\}_{i,j}
\end{equation}
where $i=0,\ldots,(I-1)L$, $j=0,\ldots,(J-1)P$, and some integers $L,P\geq1$. In the signal processing community, reconstructing a function on the domain $\mathcal{D}_o^{L,P}$ (the high resolution signal) from knowledge of the function on $\mathcal{D}_o$ (the low resolution signal) is known as super-resolution. Fig.~\ref{fig:sr_inpainting} illustrates how the different sets $\mathcal{D}_o$, $\mathcal{D}_o^{L,P}$, and $\mathcal{S}_o$ are placed under the inpainting and super-resolution framework.

In summary, we aim at designing an estimator $g_\mathbf{w}$ with the structure of a neural network where its parameters are real-valued weights $\mathbf{w}$ learned from simulated data. In particular, for a given set of frequencies of interest $\{\omega_k\}_{k=1}^K$, the estimator is defined as follows
\begin{eqnarray}
g_\mathbf{w}:&\ \mathbb{R}^{|\mathcal{S}_0|K}&\ \to\ \mathbb{R}^{|\mathcal{D}_0^{L,P}|K}\nonumber\\
&\{s(\mathbf{r},\omega_k)\}_{\mathbf{r}\in\mathcal{S}_o,k}&\ \mapsto\ \{\hat{s}(\mathbf{r},\omega_k)\}_{\mathbf{r}\in\mathcal{D}_o^{L,P},k}.
\end{eqnarray}
The goal is then that the error
\begin{equation}
\frac{\sum_{\mathbf{r}\in\mathcal{D}_o^{L,P}}|s(\mathbf{r},\omega_k)-\hat{s}(\mathbf{r},\omega_k)|^2}{\sum_{\mathbf{r}\in\mathcal{D}_o^{L,P}}|s(\mathbf{r},\omega_k)|^2}
\end{equation} 
is reduced for each frequency point. 

\begin{figure}[!t]
	\centering
	\includegraphics[scale=1.5]{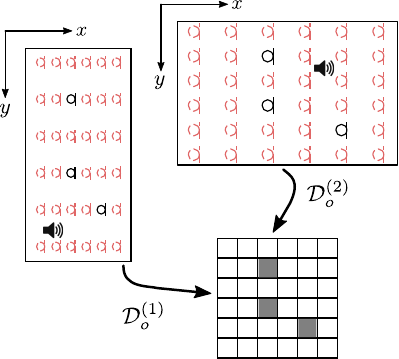}
	\caption{Example of the location agnostic property. Two rooms with different sizes lead to different rectangular grids in the Euclidean space, i.e. $\mathcal{D}_o^{(1)}\neq\mathcal{D}_o^{(2)}$. For a given frequency, we use a matrix to represent the input to the network. However, the measured and missing values in both cases (in black and red respectively) are placed at the same matrix entries. This essentially disregards any information about their locations in the Euclidean space. Similarly, the source location is considered unknown. (Color online.)}
	\label{fig:agnostic_location}
\end{figure}

It is important to note that the actual input to the neural network will represent the values $\{s(\mathbf{r},\omega_k)\}_{\mathbf{r}\in\mathcal{D}_o,k}$ in the rectangular grid $\mathcal{D}_o$ as a tensor---the missing values will be included by means of a mask on the original grid. For each frequency, this can be seen as a matrix. This implies that there is no information whatsoever provided at the input about the location of these values in the Euclidean coordinate system, i.e. the algorithm is location agnostic. In other words, irrespective of the room dimensions, we assume that our algorithm accepts measurements from a rectangular grid, whose absolute size depends on the room size. In the same way an image reconstruction algorithm would learn to recover images that have been stretched, shrunk, or zoomed in or out (see Fig.~\ref{fig:agnostic_location}). Thus, the absolute separation of points along each dimension is not the same. For example, in a room with dimensions $l_x\times l_y$, input points will be at distance of $\frac{l_x}{I}$ and $\frac{l_y}{J}$.

We will occasionally use tensors in order to represent function values on discrete spatial and frequency domains and as the data structure for the neural network operations. In particular, tensors, irrespective of their order, are denoted by bold uppercase letters, e.g. matrices can be denoted by $\mathbf{A}\in\mathbb{R}^{n_1\times n_2}$ for $n_1,n_2\in\mathbb{N}$. Regarding function values, we interchangeably use the tensor representation. For example, consider $\{s(\mathbf{r},\omega_k)\}_{\mathbf{r}\in\mathcal{D}_o^{L,P},k}$, then it possible to arrange its values into a tensor $\mathbf{S}\in\mathbb{R}^{IL\times JP\times K}$ whose elements are given by
\begin{equation}
    \mathbf{S}_{i+1,j+1,k}:=s\Big(i\frac{l_x}{(I-1)L},j\frac{l_y}{(J-1)P},\omega_k\Big).
\end{equation}

\section{\label{sec:Approach} Approach}

We propose a learning algorithm capable of estimating the magnitude of the spatial sound field, for a given frequency range, at a predefined number of locations based on very few measurements from irregularly distributed microphones. The microphones are assumed to provide the room transfer functions (RTFs) at those particular locations for a given frequency range. It is assumed that these microphones are located in a rectangular grid with a predefined number of points irrespective of the room size (see Fig.~\ref{fig:agnostic_location}). Note that the source location is also considered unknown. The prediction algorithm then provides an estimate of the corresponding RTFs at the desired locations.

\begin{figure*}[!t]
	\centering
	\includegraphics[width=\linewidth]{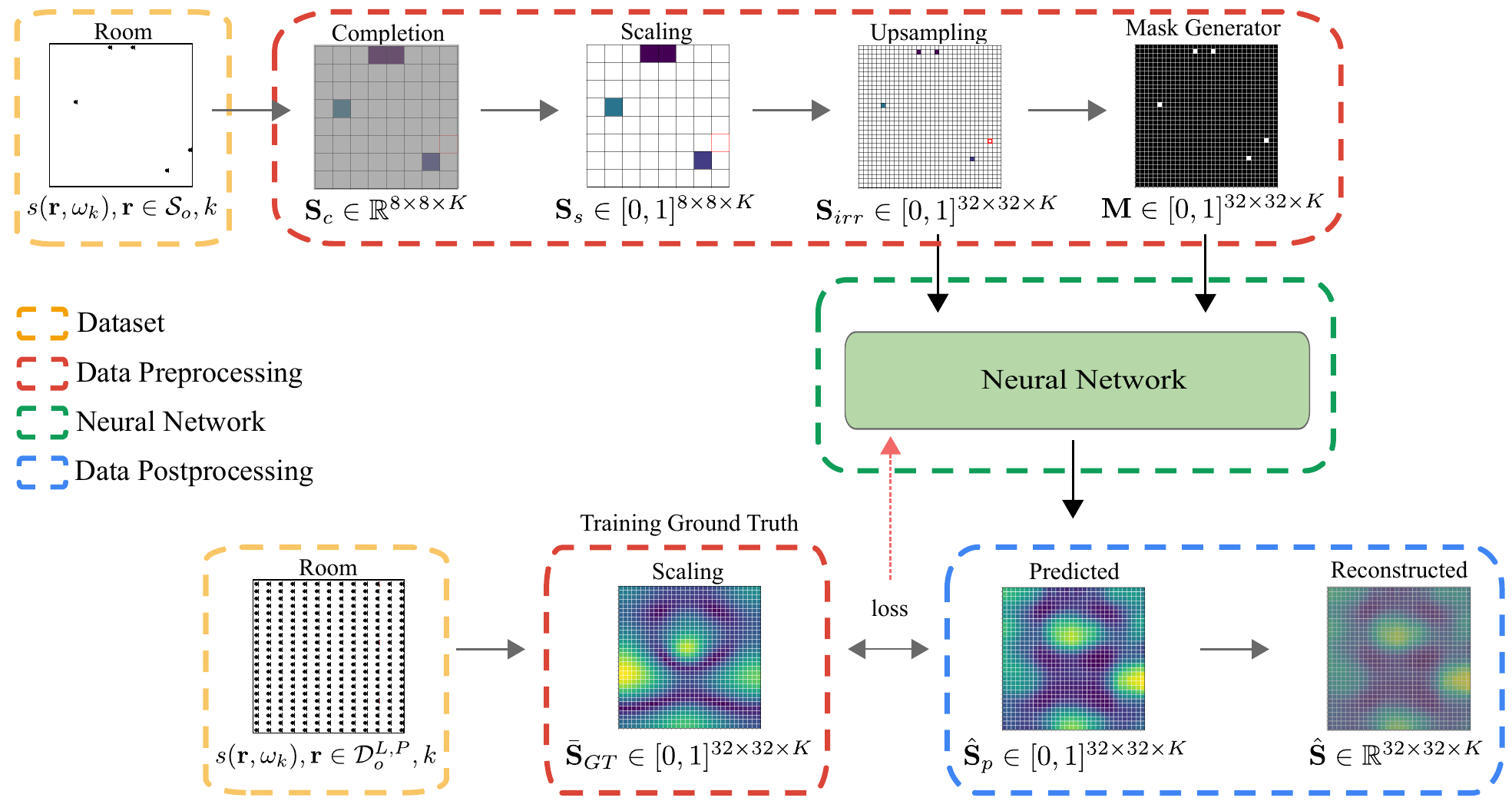}
	\caption{Diagram showing the different steps of the algorithm design. The data is assumed to be represented as third-order tensors in order to include the frequency dimension and the spatial dimensions; however, for the sake of illustration, the former is not shown. The preprocessing stage generates the input mask together with an upsampled and scaled version of the observed samples. The training examples are also scaled. For our choice of parameters, the two input tensors and the training examples take values in $[0,1]^{32\times32\times40}$. During training, the observed sample values are drawn from our simulated dataset of sound fields in rooms. (Color online.)}
	\label{fig:digram}
\end{figure*}

The approach is to train an artificial neural network that learns the structure of these sound fields from thousands of different examples of common domestic rectangular rooms. The main parts of the algorithm, which we describe in detail in the following sections, and illustrate in Fig.~\ref{fig:digram}, can be briefly summarized as follows:
\begin{itemize}
  \item \textbf{Dataset:} we simulate three-dimensional sound fields, in the frequency band [30,300] Hz, for thousands of common rectangular rooms. The magnitude of the pressure in the available spatial sample points $\mathcal{S}_o$ serves as input to the network after a preprocessing step. The magnitude of the pressure in the finer rectangular grid, i.e. $\{s(\mathbf{r},\omega_k)\}_{\mathbf{r}\in\mathcal{D}_o^{L,P}, k}$, is then used to train the network in a supervised manner.
  \item \textbf{Data Preprocessing:} from $\{s(\mathbf{r},\omega_k)\}_{\mathbf{r}\in\mathcal{S}_o,k}$, we generate a grid version, defined on $\mathcal{D}_o^{L,P}$, consisting of the observed samples and a mask that encodes the information about the locations of these measurements. This preprocessing step involves completion, scaling, and upsampling operations.
  \item \textbf{Neural Network:} The architecture learns to predict a scaled version of the two-dimensional function $\{s(\mathbf{r},\omega_k)\}_{\mathbf{r}\in\mathcal{D}_o^{L,P},k}$  from the preprocessed observed sample values $\{s(\mathbf{r},\omega_k)\}_{\mathbf{r}\in\mathcal{S}_o,k}$ and the mask.
  \item \textbf{Data Postprocessing:} Estimates the appropriate scaling in order to restore the predicted values to the range of the source data.
\end{itemize}

The data and code of the proposed algorithm is freely available online \footnote{See supplementary material at \url{github.com/francesclluis/sound-field-neural-network}}.

\subsection{\label{subsec:dataset}Dataset}

The sound field in a lightly damped rectangular room can be approximated using Green's function expressing the solution as an infinite summation of room modes (or standing waves) in the x-,y-, and z-dimension of the room \cite{jacobsen2013fundamentals}
\begin{equation}
    G(\mathbf r,\mathbf r_0, w) \approx -\frac{1}{V}\sum_{N} \frac{\psi_{N}(\mathbf r) \psi_{N}(\mathbf r_0)}{(\omega/c)^2 - (\omega_{N}/c)^2 -j\omega /\tau_N}.  
\end{equation}
Here, for compactness $\sum_N$ denotes a triple summation across the modal order in each dimension of the room i.e. $\sum_N = \sum_{n_x=0}^\infty \sum_{n_y=0}^\infty \sum_{n_z=0}^\infty$ and correspondingly $N$ represents the triplet of integers ${n_x,n_y,n_z}$. The volume of the room is denoted $V$, $\psi_N(\cdot)$ is the mode shape associated with a specific $N$, $\omega_{N}$ is the angular resonance frequency of the mode, $\tau_N$ is the time constant of the mode, and $c$ is the speed of sound. The room shape is here determined assuming rigid boundaries leading to the expression
\begin{equation}
    \psi_N(\mathbf x) = \Lambda_N \cos\left( \frac{n_x \pi x}{l_x} \right) \cos\left( \frac{n_y \pi y}{l_y} \right) \cos\left( \frac{n_z \pi z}{l_z} \right),
\end{equation}
where $\Lambda_N = \sqrt{\epsilon_{n_x} \epsilon_{n_y} \epsilon_{n_z}}$ are normalization constants with $\epsilon_0 = 1$, $\epsilon_1 = \epsilon_2 = \ldots = 2$.

Throughout this work, the focus is to predict the variation of the sound field in a single $xy$-plane, hence, we seek to train a model which can predict the variation of the sound pressure in the plane. With the purpose to generalize for any $xy$-plane, we remove the height variation in the Dataset by setting $n_z = 0$. The time constants of each mode are determined from the absorption coefficient calculated using Sabine's equation and assuming a reverberation time $T_{60}$ of 0.6~s and uniform distribution of absorption on the surfaces of the room.

We use this model to simulate point source radiation in 5~000 rectangular rooms. Room size and room proportions are randomly created following the recommendation for listening room dimensions for audio reproduction in the standard ITU - R BS.1116 - 3 \cite{international2015recommendation}. The floor area ranges from \SIrange{20}{60}{\metre\squared} and the dimension ratios follow:
\begin{equation}
1.1\frac{l_y}{l_z} \leq \frac{l_x}{l_z} \leq 4.5 \frac{l_y}{l_z} -4
\end{equation}
where $l_x$, $l_y$, and $l_z$ correspond to length, width, and height respectively. In addition, the source is placed at a random $xy$-location, i.e. $(x_o, y_o, 0)$ for $x_o\in(0,l_x)$ and $y_o\in(0,l_y)$. Both the dimensions and source location are sampled uniformly.

The magnitude of the sound field pressure is acquired in the finer rectangular grid $\mathcal{D}_o^{L,P}$ with $L=P=4$ and $I=J=8$. This essentially divides the room into a grid of 32 by 32 uniformly-spaced points independently of its dimensions.  We analyze the results with 1/12th octave frequency resolution in the range [30, 300]~Hz including all room modes with a resonance frequency below 400~Hz. This gives $K=40$ frequency points. The sound fields generated using this technique are referred to as ground truth sound fields, i.e. $s_{GT}(\mathbf{r}, \omega_k):=s(\mathbf{r}, \omega_k)$ for $\mathbf{r}\in\mathcal{D}_o^{L,P}$ and $k=1,\ldots K$. A subset of $s_{GT}(\mathbf{r}, \omega_k)$ containing the observed samples captured by the microphones, $\{s_{GT}(\mathbf{r},\omega_k)\}_{\mathbf{r}\in\mathcal{S}_o,k}$, is used in the preprocessing part.

\subsection{Preprocessing}

This part addresses the processing stage necessary to handle the arbitrary nature of the sampling distribution. In particular, the raw input data is allowed to be variable in size and sampling location. In order to address this, we complete the input data to take values on $\mathcal{D}_o$. This is followed by a scaling operation in order to generalize the predictions for arbitrary sources and receivers. The actual information of where the samples are located within $\mathcal{D}_o^{L,P}$ is encoded into a mask-like function. An upsampled version of this processed input data together with this mask comprises the final input to the network.

\subsubsection{Completion}

We assume that the possible observed pressure values correspond to locations within the coarser grid $\mathcal{D}_o$, which also covers the whole room area. In this paper, the choice of parameters results in $\mathcal{D}_o$ being a grid of $8$ by $8$ points. The samples observed are then given by $\{s_{GT}(\mathbf{r}, \omega_k)\}_{\mathbf{r}\in\mathcal{S}_o,k}$. Irrespective of the structure of $\mathcal{S}_o$, i.e. the number and pattern of observed samples, the neural network is designed so that the size of the input data is fixed. In order to address this, we introduce a function defined on $\mathcal{D}_o$ that, in a sense, completes the acquired data, i.e.
\begin{equation}
s_{c}(\mathbf{r}, \omega_k) :=
\left\{
	\begin{array}{ll}
		s_{GT}(\mathbf{r},\omega_k)  & \mbox{if } \mathbf{r}\in\mathcal{S}_o\\
		\max_{\mathbf{r^\prime}\in\mathcal{S}_o} s_{GT}(\mathbf{r^\prime},\omega_k) & \mbox{if } \mathbf{r}\not\in\mathcal{S}_o.
	\end{array}
\right.
\end{equation}
for each $\omega_k$. In other words, for the locations where no samples are provided, i.e. no microphone is present, $s_{c}$ is chosen arbitrarily to take the maximum value.

\subsubsection{Scaling}

We want the proposed method to be independent of the gain in the measurement equipment and the reproduction system. Thus, we introduce a scaling for the sample values $s_{c}$ in such a way that the range is restricted to [0,1], i.e.
\begin{equation}
s_{s}(\mathbf{r},\omega_k):=\frac{s_{c}(\mathbf{r},\omega_k)-\min_{\mathbf{r}\in\mathcal{S}_o}s_{c}(\mathbf{r},\omega_k)}{\max_{\mathbf{r}\in\mathcal{S}_o}s_{c}(\mathbf{r},\omega_k) - \min_{\mathbf{r}\in\mathcal{S}_o}s_{c}(\mathbf{r},\omega_k)}
\end{equation}
for each $\omega_k$. Consequently, the neural network will learn to predict the sound field values in [0,1]. A postprocessing stage will be added so that the predictions are restored to the original range.

\subsubsection{Upsampling}

Since we are interested in predicting values in the finer rectangular grid, $\mathcal{D}_o^{L,P}$, we transform $s_{s}\in\mathbb{R}^{8\times8\times40}$ to a function $s_{irr}\in\mathbb{R}^{32\times32\times40}$ by means of an upsampling operation. This new function $s_{irr}$ consists of a scaled version of the irregularly-distributed microphone measurements. In particular, we have that
\begin{equation}
s_{irr}(\mathbf{r}, \omega_k):=
\left\{
	\begin{array}{ll}
		s_{s}(\mathbf{r},\omega_k)  & \mbox{if } \mathbf{r}\in\mathcal{D}_o\\
		1 & \mbox{if } \mathbf{r}\in\mathcal{D}_o^{L,P}\setminus\mathcal{D}_o
	\end{array}
\right.
\end{equation}
for each $\omega_k$. The original measurements are incorporated into $s_c$, however, the actual input values to the network are given by $s_{irr}$. Note that the value of $s_{irr}$ for $\mathbf{r}\in\mathcal{D}_o^{L,P}\setminus\mathcal{D}_o$ can be arbitrarily chosen due to the mask-related operation that follows.

\subsubsection{Mask generator}

The function $s_{irr}$ does not provide any information about which values have been originally observed. Thus, we simultaneously generate a mask, defined on the finer grid $\mathcal{D}_o^{L, P}$, that carries information about the spatial locations of the measurements. This mask takes the value 1 at each available spatial sample point and 0 otherwise, i.e.
\begin{equation}
m(\mathbf{r},\omega_k):=
\left\{
	\begin{array}{ll}
		1  & \mbox{if } \mathbf{r}\in\mathcal{S}_o \\
		0 & \mbox{if } \mathbf{r}\in\mathcal{D}_o^{L, P}\setminus\mathcal{S}_o
	\end{array}
\right.
\end{equation}
for all $\omega_k$. Clearly, the mask must be the same for every frequency point.

\subsubsection{Input}
The input data to the network consists of third-order tensors representing the frequency dimension and the two spatial dimensions, i.e. $\textbf{M}\in[0,1]^{32\times32\times40}$ and $\textbf{S}_{irr}\in[0,1]^{32\times32\times40}$. It is important to emphasize that the network performs convolutions considering the three dimensions in order to learn the relationships within and between frequency and space.

\subsection{Neural Network}
\subsubsection{Architecture}
\begin{figure}[!t]
	\centering
	\includegraphics[width=\columnwidth]{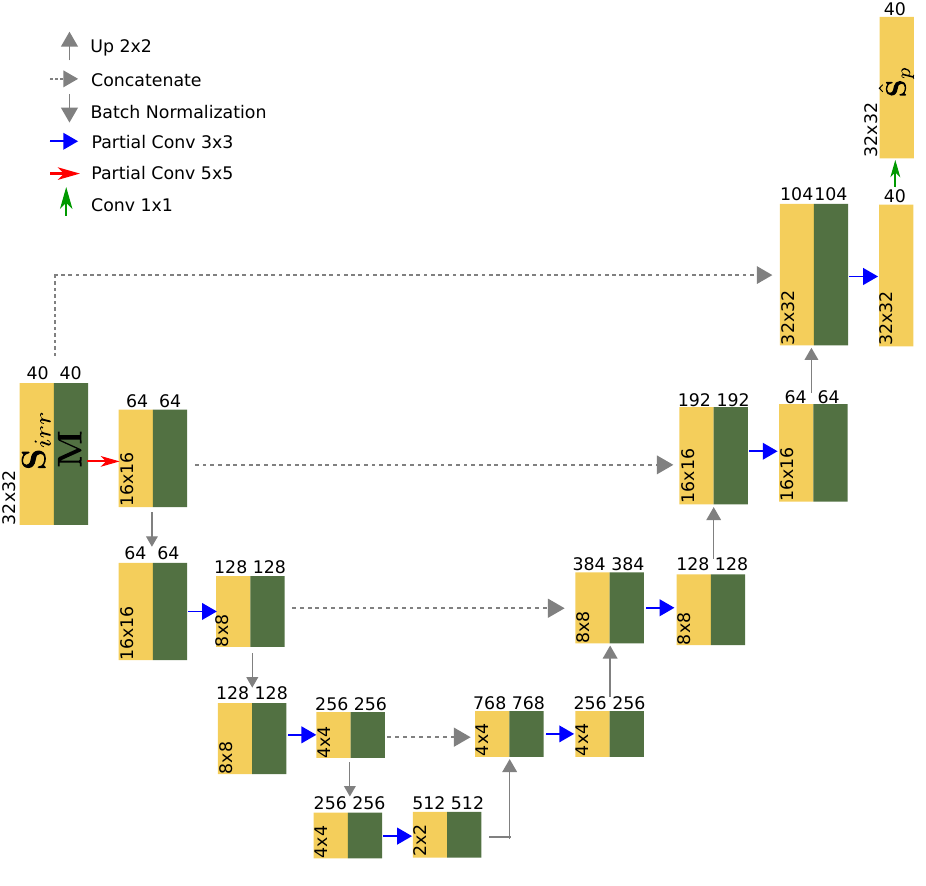}
	\caption{Schematic diagram of the neural network architecture proposed in this paper. This diagram is not exhaustive in terms of all the operations involved. For further details, the reader can refer to the text. (Color online.)}
	\label{fig:neural-network}
\end{figure}

We propose a U-Net-like deep neural network \cite{Ronneberger:2015aa} with partial convolutions \cite{Liu:2018aa} in order to predict the magnitude of the sound field pressure in a room. U-Net was first introduced for the task of biomedical image segmentation and since then it has been successfully used in many cases.

The U-Net encoder-decoder structure can learn multi-resolution features of the sound field in the frequency-space domain, i.e. it can capture the sound field variations at different scales in both domains. This is carried out by the encoder which halves the feature maps by using a stride of 2 and doubling the filter size in each partial convolution. The decoder then reverses this procedure by upsampling the feature maps and reducing by 2 the filter size. After each partial convolution, the encoder uses a ReLU activation while the decoder uses a Leaky ReLU activation with a negative slope coefficient of 0.2. Furthermore, the decoder, through concatenation, incorporates at the same hierarchical level the feature maps and masks computed by the encoder. In other words, the features from different resolutions in the frequency-space domain are also utilized as an input in the upsampling layers of the decoder. Finally, a 1$\times$1 convolution with a sigmoid activation projects the last feature map to generate the predicted sound field $\hat{\mathbf{S}}_p$. Fig.~\ref{fig:neural-network} shows a schematic diagram of the architecture.

Although there are similarities between U-Net and a standard encoder-decoder architecture, their skip connections are paramount in order to attain better performance. This has been shown by ablation studies in image segmentation \cite{drozdzal2016importance} and label-to-image\cite{isola2017image} tasks. Skip connections allow U-Net to access low-level information that may be lost when propagated through the network. In the current case, skip connections help to recover spatial information lost during downsampling which corresponds to the initial arrangement of measurements.

\subsubsection{Partial Convolutions}
Unlike traditional convolutions, partial convolutions\cite{Liu:2018aa} allow us to compute the output feature maps based solely on the available spatial sample points from the input feature maps. This provides the necessary flexibility to use any number of microphones at irregularly distributed locations. Let $w$ be the sliding convolutional window with size $k_h\times k_t$. Consider further $\mathbf{I}_w\in\mathbb{R}^{k_h\times k_t\times C}$ and $\mathbf{M}_w\in[0, 1]^{k_h\times k_t\times C}$ as corresponding to the $C$-channel input feature maps and mask within $w$ respectively. The tensor $\mathbf{W}\in\mathbb{R}^{k_h\times k_t\times C'\times C}$ respresents the filter weights and $\mathbf{b}\in\mathbb{R}^{C'}$ is the bias. Partial convolution computes each spatial location value $\mathbf{o}_w'\in\mathbb{R}^{C'}$ in the $C'$-channel  output feature maps as
\begin{equation}
\mathbf{o}_w' :=
\left\{
	\begin{array}{ll}
		\mathbf{W}\cdot (\mathbf{I}_{w}\odot \mathbf{M}_{w})\frac{\textrm{sum}(\mathbf{1})}{\textrm{sum}(\mathbf{M}_w)}+\mathbf{b}  & \mbox{if } \textrm{sum}(\mathbf{M}_w) > 0 \\
		0 & \mbox{otherwise}
	\end{array}
\right.
\end{equation}
where $\mathrm{sum}(\cdot)$ receives a tensor as an argument and provides the summation of its elements, $\odot$ is the Hadamard product, and $\cdot$ is a combination, in different dimensions, of matrix dot products and element-wise summations\cite{Liu:2018aa}. The scaling factor $\frac{\textrm{sum}(\mathbf{1})}{\textrm{sum}(\mathbf{M}_w)}$ can be interpreted as a measure of the amount of known information in the input feature maps. Then, the mask $\mathbf{M}_w$ is updated at each spatial location $\mathbf{m}'\in\mathbb{R}^{C'}$ as follows:
\begin{equation}
\mathbf{m}' =
\left\{
	\begin{array}{ll}
		1 & \mbox{if } \textrm{sum}(\mathbf{M}_w) > 0 \\
		0 & \mbox{otherwise}.
	\end{array}
\right.
\end{equation}

\subsubsection{\label{subsec:LossFunction}Loss Function}
In order to train the model in a supervised manner, we also use a scaled version of the ground truth in order to be consistent with the output data before postprocessing. The assumption is that this process may also assist the learning process. The scaling is given by
\begin{equation}
\bar{s}_{GT}:=\frac{s_{GT}(\mathbf{r},\omega_k)-\min s_{GT}(\mathbf{r},\omega_k)}{\max s_{GT}(\mathbf{r},\omega_k) - \min s_{GT}(\mathbf{r},\omega_k)}
\end{equation}
for $\mathbf{r}\in\mathcal{D}_o^{L,P}$ and $k=1,\ldots,K$. It is clear then that $\bar{s}_{GT}(\mathbf{r},\omega_k)\in[0,1]$.

As a loss function, we use two terms in order to distinguish between predicted values in the available spatial sample points $\mathcal{S}_o$ and its complement under $\mathcal{D}_o^{L,P}$. We first define
\begin{equation}
\mathcal{L}_{\mathcal{S}_o} := \frac{\mathrm{sum}\big(\big\lvert \mathbf{M} \odot (\hat{\mathbf{S}}_p - \bar{\mathbf{S}}\big)\big\rvert\big)}{IL\times JP\times K}
\label{eq:av_loss}
\end{equation}
and then
\begin{equation}
\mathcal{L}_{D_o^{L,P}\setminus\mathcal{S}_o} := \frac{\mathrm{sum}\big(\big\lvert(\mathbf{1}-\mathbf{M}) \odot (\hat{\mathbf{S}}_p - \bar{\mathbf{S}}_{GT})\big\rvert\big)}{IL\times JP\times K}
\label{eq:no_av_loss}
\end{equation}
where $\mathbf{1}\in\mathbb{R}^{32\times32\times40}$ with all entries equal to 1, and sum($|\cdot|$) acting on a tensor is the summation of the absolute value of its elements. The combined loss function finally takes the form
\begin{equation}
\mathcal{L}:= \mathcal{L}_{\mathcal{S}_o} + 12\mathcal{L}_{D_o^{L,P}\setminus\mathcal{S}_o}.
\label{eq:total_loss}
\end{equation}
The factors in (\ref{eq:total_loss}) were chosen as the best performing ones after analyzing the performance on 1~000 validation rooms.

\subsubsection{Training Procedure}
The model is trained in two different stages using supervised learning. We use 75\% of the dataset for training purposes and the remaining 25\% is used for validation. For both stages, the model is trained during 400 epochs and the weights with less validation loss are selected. In the first stage, the learning rate is set to $2\cdot 10^{-4}$ and batch normalization is enabled in all layers. For the second stage, the learning rate is set to $5\cdot 10^{-5}$ with batch normalization disabled in all encoding layers. Training the model in multiple stages helps to overcome the error generated by batch normalization when computing, in the first stage, the mean and variance for all input values, corresponding to known and unknown locations. In addition, faster convergence is achieved.

\subsection{\label{subsec:postprocessing}Postprocessing}
We use linear regression to restore the output of the neural network $\hat{s}_p$ to its original range. Thus, the rescaled version takes the form
\begin{equation}
\hat{s}(\mathbf{r}, \omega_k)= a_k \cdot \hat{s}_p(\mathbf{r}, \omega_k) + b_k
\end{equation}
for all $\mathbf{r}\in\mathcal{D}_o^{L,P}$ and $k=1,\ldots,K$, where the values $a_k,b_k\in\mathbb{R}$ are determined through the following optimization problem
\begin{equation}
\min_{a_k,b_k\in\mathbb{R}}\sum_{\mathbf{r}\in\mathcal{S}_o}|a_k\cdot \hat{s}_p(\mathbf{r}, \omega_k) + b_k-s_{c}(\mathbf{r}, \omega_k)|^{2}
\end{equation}
for each $k=1,\ldots,K$. Note that the rescaling operation could be implemented as another neural network that learns the mapping function. However, experiments showed that linear regression provided reasonable performance.

\section{Results}
\label{sec:Results}
\subsection{Evaluation Metrics}
We use two different measures of performance for the proposed method. First, we consider the normalized mean square error (NMSE) computed for each frequency point, i.e. 
\begin{equation}
    \mathrm{NMSE}_k = \frac{\sum_{\mathbf{r}\in\mathcal{D}_o^{L,P}}|s(\mathbf{r},\omega_k)-\hat{s}(\mathbf{r},\omega_k)|^2}{\sum_{\mathbf{r}\in\mathcal{D}_o^{L,P}}|s(\mathbf{r},\omega_k)|^2}.
\end{equation}
The NMSE mainly provides an average absolute squared error over all locations between the reconstructed and the original signals. As a consequence, a high NMSE value may result from a poor performance locally while performing individually well in the remaining spatial locations.

Therefore, we use the concept of mean structural similarity\cite{wang2004image} (MSSIM) from image processing. This evaluates how the model predicts the overall shape of the pressure distribution for each frequency point. Moreover, it also provides a measure of performance that is independent of the scaling chosen. Let us first introduce the structural similarity index (SSIM) between two matrices $\mathbf{A}, \mathbf{B}\in\mathbb{R}^{n\times n}$ as follows
\begin{equation}
\mathrm{SSIM}(\mathbf{A}, \mathbf{B}) = \frac{(2\mu_{\mathbf{A}}\mu_{\mathbf{B}}+c_{1})(2\sigma_{\mathbf{A}\mathbf{B}}+c_{2})}{(\mu_{\mathbf{A}}^{2}+\mu_{\mathbf{B}}^{2}+c_1)(\sigma_{\mathbf{A}}^{2}+\sigma_{\mathbf{B}}^{2}+c_{2})}
\end{equation}
where $\mu$ is the mean of the corresponding matrix entries, $\sigma^2$ the estimate of the variance of the entries, and $\sigma_{\mathbf{A}\mathbf{B}}$ is the covariance estimate between the entries of $\mathbf{A}$ and $\mathbf{B}$. The constants $c_{1}=(h_{1}R)^2$ and $c_{2}=(h_{2}R)^2$, where $R$ is the dynamic range of the entry values, are meant to stabilize the division with a weak denominator. We set $h_{1}$ and $h_{2}$ to the standard values 0.01 and 0.03 respectively.

In our scenario, we consider the individual matrices $\mathbf{S}_k\in\mathbb{R}^{IL\times JP}$, i.e. the $k$-th matrix of tensor $\mathbf{S}\in\mathbb{R}^{IL\times JP\times K}$. Now, let $\{\mathbf{S}^n_k(\eta)\}_{n=1}^N$ denote the set of all possible windowed versions of $\mathbf{S}_k$ of size $\eta\times\eta$. The mean structural similarity is then given by
\begin{equation}
    \mathrm{MSSIM}(\mathbf{S}_k, \hat{\mathbf{S}}_k) := \frac{1}{N} \sum_{n=1}^{N}\mathrm{SSIM}(\mathbf{S}^n_k(\eta), \hat{\mathbf{S}}^n_k(\eta))
\end{equation}
for each frequency point. In the results presented, we have used $\eta=7$.

\subsection{\label{sec:3} Simulated Data}
\begin{figure}[!t]
	\centering
	\includegraphics{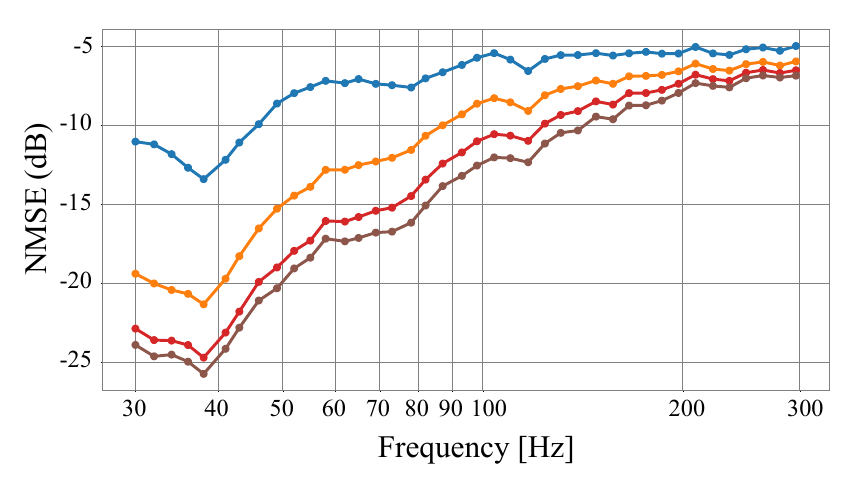}
	\caption[]{Normalized mean squared error (NMSE) estimated from simulated data. The results are reported for different number of microphone observations $n_{mic}$, i.e. (\bluelegend):$n_{mic} = 5$, (\orangelegend):$n_{mic} = 15$, (\redlegend):$n_{mic} = 35$, and (\brownlegend):$n_{mic} = 55$. (Color online.)}
	\label{fig:NMSE_sim}
\end{figure}

\begin{figure}[!t]
	\centering
	\includegraphics{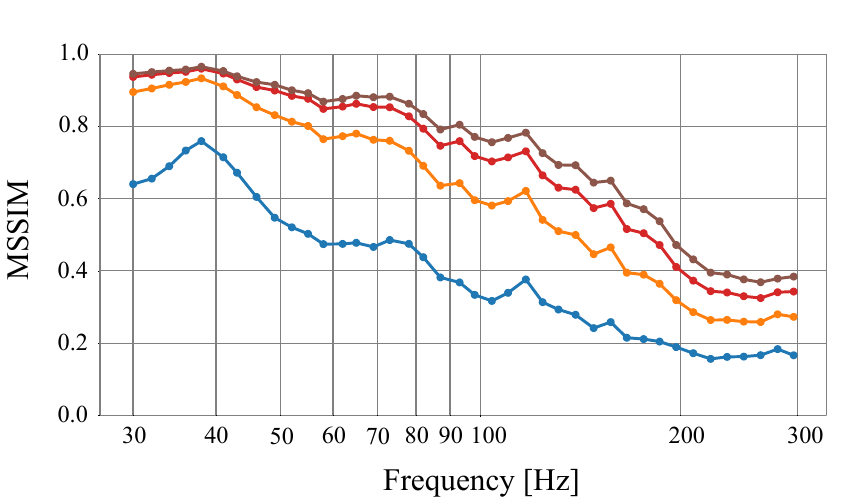}
	\caption[]{Mean structural similarity index (MSSIM) estimated from simulated data. The results are reported for different number of microphone observations $n_{mic}$, i.e. (\bluelegend):$n_{mic} = 5$, (\orangelegend):$n_{mic} = 15$, (\redlegend):$n_{mic} = 35$, and (\brownlegend):$n_{mic} = 55$. (Color online.)}
	\label{fig:SSIM_sim}
\end{figure}

We asses the reconstruction performance of the proposed method, i.e. the generalization error, by using sound fields in 30 different rooms. These have been simulated identically to the training data and have not been previously seen by the network. We are interested in evaluating the performance with regard to the number of irregularly placed microphones, denoted by $n_{mic}$. Thus, given $n_{mic}$, we analyze the reconstruction in each room placing the microphones in 10~000  different arrangements, i.e. each realization corresponds to a different $\mathcal{S}_o$. Figures \ref{fig:NMSE_sim} and \ref{fig:SSIM_sim} show, as a function of frequency, the average NMSE in dB and MSSIM for all rooms and locations tested and different number of available microphones. 

Results show a general improved performance in sound field reconstruction as the number of available microphones is increased. At the same time, performance degrades as the frequency increases. This is in agreement with theoretical results that, given a maximum frequency content, require a higher sampling density for a more robust reconstruction and, given a reconstruction error, the sampling density constraints also increase whenever higher frequency content is available\cite{Landau:1967aa,Ajdler:2006aa}. This suggests that the neural network capacity is subject to the same physical limitations as classical methods when learning the spatial variations of the pressure distribution. In other words, at high frequencies it is hindered by undersampling and also requires more observations to improve robustness. For example, the relative improvement as the number of microphones increase is higher at lower frequencies as opposed to the high-frequency range. It is at this high frequency range where more observations do not provide a big impact on performance. However, the requirements in terms of sampling density for a particular performance seem to be less stringent than other methods present in the literature. For example, only $n_{mic}=5$ microphones are able to provide an NMSE below $-5$~dB for the frequency range considered in common domestic rooms.

It is also important to observe that the loss functions defined in Eq.~\ref{eq:av_loss} and Eq.~\ref{eq:no_av_loss} are suitable for prediction at low frequencies but they underperform at high frequencies. These commonly result in predictions that emphasize the median value in order to reduce the overall error. This can explain, in the frequency range 100-300~Hz, the more abrupt changes in performance of the MSSIM as opposed to the NMSE.

\subsection{\label{sec:4} Experimental Data}

\begin{figure}[!t]
	\centering
	\includegraphics{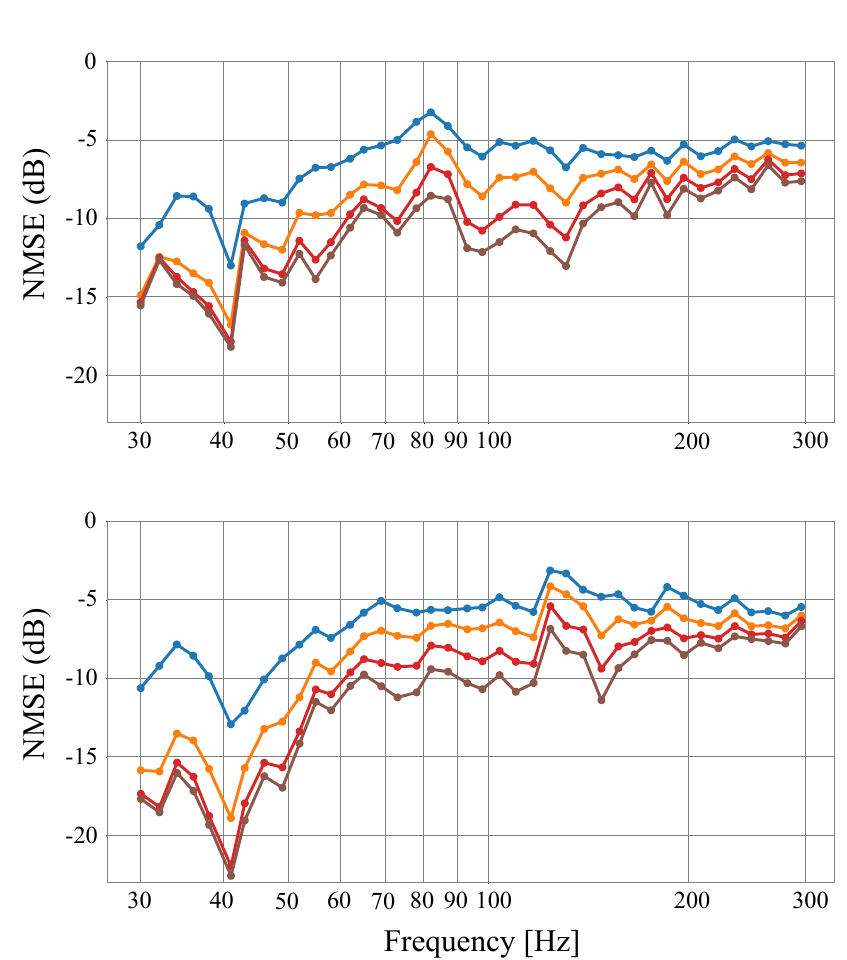}
	\caption[]{Normalized mean square error (NMSE) in dB estimated from experimental data. Top and bottom plots correspond to different source locations. The results are reported for different number of microphone observations $n_{mic}$, i.e. (\bluelegend):$n_{mic} = 5$, (\orangelegend):$n_{mic} = 15$, (\redlegend):$n_{mic} = 35$, and (\brownlegend):$n_{mic} = 55$. (Color online.)}
	\label{fig:NMSE_real}
\end{figure}

\begin{figure}[!t]
	\centering
	\includegraphics{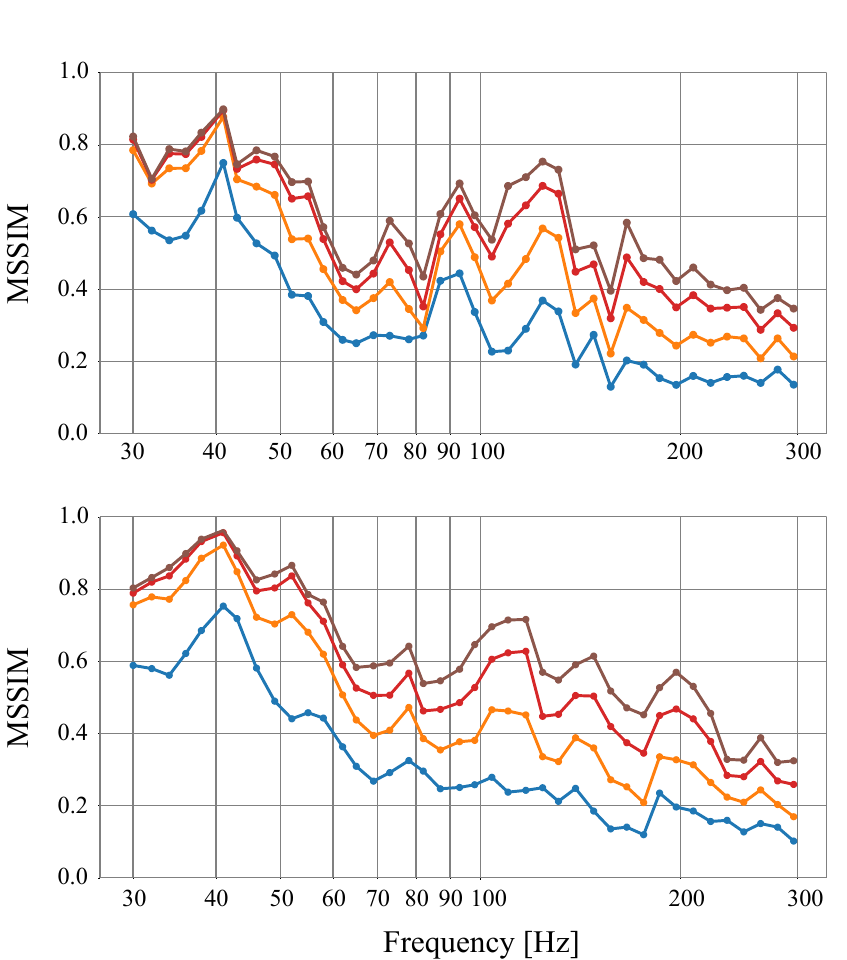}
	\caption[]{Mean structural similarity (MSSIM) estimated from experimental data. Top and bottom plots correspond to different source locations. The results are reported for different number of microphone observations $n_{mic}$, i.e. (\bluelegend):$n_{mic} = 5$, (\orangelegend):$n_{mic} = 15$, (\redlegend):$n_{mic} = 35$, and (\brownlegend):$n_{mic} = 55$. (Color online.)}
	\label{fig:SSIM_real}
\end{figure}

We test the model optimized for simulated data in a real listening room. The RTFs are estimated for two different source locations on a two-dimensional grid consisting of 32 by 32 points uniformly spaced along the corresponding dimensions. In particular, impulse response measurements were conducted from two 10'' loudspeakers on a grid one meter above the floor in a rectangular room of dimensions $4.16\times6.46\times 2.3$~m. The measurements were performed using 4-second duration exponential sweeps from 0.1~Hz to 24~kHz at a sampling frequency of 48~kHz \cite{farina2000simultaneous}. These measurements were performed with two microphones, each covering roughly half of the grid. The microphones were a Br\"{u}el \& Kj\ae r (B\&K) 4192 and a B\&K 4133 $\frac{1}{2}$'' condenser microphone connected to a B\&K Nexus conditioning amplifier and recorded with an RME Fireface UFX+ sound card. Both microphones were level calibrated at 1~kHz using a B\&K 4231 calibrator prior to the measurements. The reverberation time of the room, specified as the arithmetic average of the 1/3 octave $T_{20}$ estimates\cite{iso2008acoustics} in the range of 32~Hz to 316~Hz, was 0.46~s.

Similar to the previous scenario, we investigate the performance of the model with regard to the number of microphones placed in the room. We are particularly interested in assessing the performance when using very few observations. Thus, for each predefined source location, we also use here 5, 15, 35, and 55 microphones in 10~000 different arrangements and analyze the mean performance with a 95\% confidence interval. These results are reported in Figures \ref{fig:NMSE_real} and \ref{fig:SSIM_real}.

\begin{figure*}[!t]
	\centering
	\includegraphics[width=\linewidth]{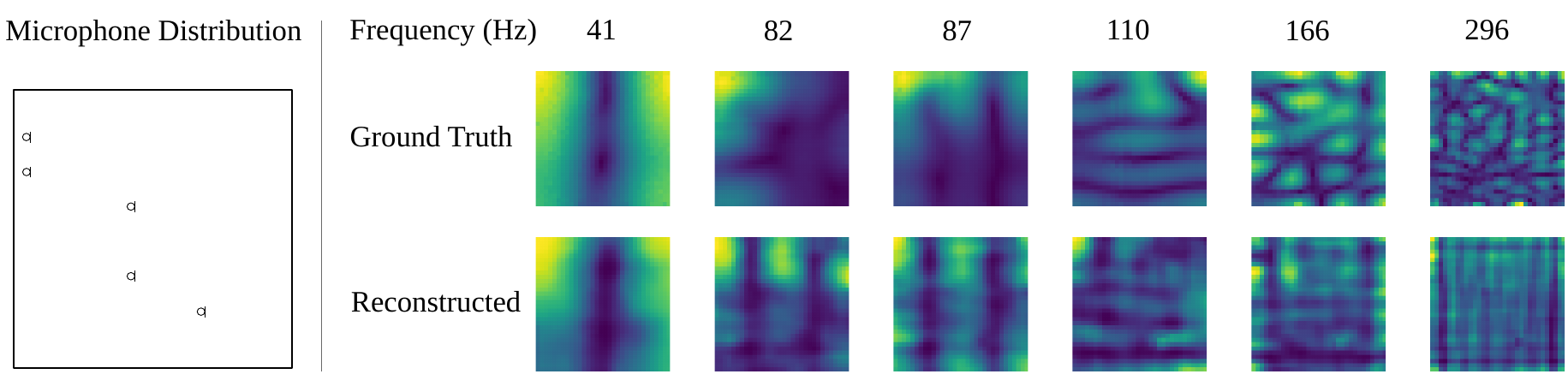}
	\caption{Visualization of the model reconstruction when using 5 microphones arbitrarily placed. The results are shown for different frequencies in a real room where the source location is the same as the top plots in Figures \ref{fig:NMSE_real} and \ref{fig:SSIM_real}. (Color online.)}
	\label{fig:visualization_predictions}
\end{figure*}

It is important to emphasize that the model was trained using simulated data. Moreover, the simulations were simplified by assuming mode shapes equal to rigid walls and removing all room modes including height variation, neither of which is true for the experimental data.
It can be observed that, given $n_{mic}$, the NMSE improves for decreasing frequencies as a general trend although there exist inconsistencies at a local level, i.e. adjacent frequencies may present abrupt changes in performance. The same interpretation applies to the MSSIM. In particular, there are two specific frequencies acting as outliers, i.e. 82~Hz and 157~Hz for the two different source locations. In this case, this is likely to be caused by the sources being positioned at nulls of the room modes. Fig.~\ref{fig:visualization_predictions} depicts a representation of the magnitude of the sound field when the reconstruction is performed using only 5 microphones.

\subsection{Computational Complexity}
Apart from the reduced number of microphones used, another advantage of the proposed method is the computational complexity regarding the inference operation. The training stage is usually time consuming, but it can often be run offline. The model size is relatively small with 3.9 million parameters resulting in a deterministic inference time of approximately 0.05~s on a Nvidia GeForce GTX 1080 Ti GPU (value estimated from 100 different room predictions).

\subsection{\label{subsec:microphone_distribution} Microphone Distribution}
\begin{figure}[!t]
	\centering
	\includegraphics[]{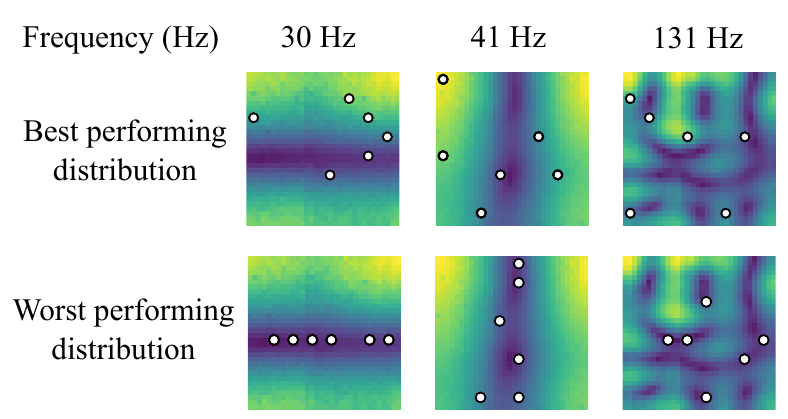}
	\caption[]{Best and worst performing sampling distributions for 6 microphones in terms of NMSE performance. The results are shown for different frequencies in a real room where the source location is the same as the top plots in Figures \ref{fig:NMSE_real} and \ref{fig:SSIM_real}. Symbol (\(\circ\)) represents the microphone locations. (Color online.)}
	\label{fig:mic-dist}
\end{figure}

In our analysis, we have mainly focused on the performance based on the number of observations. However, we are also interested in studying the impact that particular microphone distributions have on the performance. Fig.~\ref{fig:mic-dist} shows an illustration of the best and worst performing microphone distributions in terms of the NMSE. It can be observed that a better reconstruction at a specific frequency is achieved when the microphones capture the maximum variation of the pressure values. On the contrary, if the observations consist solely of the dip-like part of the room modes, the reconstruction degrades significantly. Evidently, this effect is frequency dependent, thus there is not a microphone setup that performs well across all frequencies. However, this also suggests that an unstructured microphone arrangement may be more likely to avoid these sampling issues caused by the modal structure.

\section{Discussion}
The work presented here indicates the potential for applying neural networks to predict sound field variations in an entire room from few microphone observations in a location agnostic manner. The training data was based on simplifying assumptions e.g. near-rigid walls, no room modes with height variation, and perfectly rectangular room shape. Despite the mismatch between the training and test scenarios, the network shows promising results under unseen data. This can be understood in relation to the literature where it has been shown that the structure of convolutional neural networks represents a prior which conditions the network to perform well for image-like signals\cite{Ulyanov:2018aa}. The magnitude of the spatial sound field naturally fits the latter. Further, the results can be interpreted as a transfer learning\cite{Goodfellow-et-al-2016} approach where the architecture itself helps to generalize well in the experimental scenario from weights only learned with simulated data.

Despite the discrepancy between training data and the experimental measurements, the extrapolation results are encouraging. It is, therefore, expected that the method could be extended to more complicated scenarios like non-rectangular rooms with complex boundary conditions given the appropriate training data.

\section{\label{sec:5}Conclusions}

In this paper, a deep-learning-based method for sound field reconstruction in rectangular rooms has been proposed and examined. The method jointly performs inpainting and super-resolution in order to reconstruct the magnitude of the sound pressure in a two-dimensional plane of a three-dimensional room.  The focus of this work is threefold: use a very low number of microphones, accommodate irregular and location agnostic microphone distributions, and carry out inference that is computationally efficient. The results suggest a performance which offers advantages in these three directions, e.g. even using 5 microphones arbitrarily placed the method provides an acceptable reconstruction error with a low inference time.

Regarding future work, the study of generative adversarial networks as discriminators may help to increase the performance at high frequencies. In addition, using more complex acoustic simulation models during the training stage could overcome performance inconsistencies at a local level as well as providing a lower generalization error when using experimental data.

\begin{acknowledgments}
This project has received funding from the European Union's Horizon 2020 research and innovation programme under the Marie Sk\l{}odowska-Curie grant agreement No 812719.
\end{acknowledgments}






\bibliography{sampbib}

\begin{thebibliography}{10}
\def\enquote#1,{``#1,''}
\def\enxquote#1{``#1''}
\expandafter\ifx\csname url\endcsname\relax
  \def\url#1{\texttt{#1}}\fi
\expandafter\ifx\csname urlprefix\endcsname\relax\def\urlprefix{URL }\fi
\providecommand{\bibinfo}[2]{#2}
\def\plainquote#1{``#1''}
\providecommand{\noopsort}[1]{}
\providecommand{\switchargs}[2]{#2#1}
\providecommand{\dourl}[1]{\href{http://#1}{\nolinkurl{#1}}}
  \def\eatspace #1{#1}

\bibitem{Tylka:2015aa}
\bibinfo{author}{J.~G. Tylka} and \bibinfo{author}{E.~Choueiri},
  \enquote{\bibinfo{title}{Comparison of techniques for binaural navigation of
  higher-order ambisonic soundfields}}, in \emph{\bibinfo{booktitle}{Audio
  Engineering Society Convention 139}}, \bibinfo{organization}{Audio
  Engineering Society} (\bibinfo{year}{2015}).

\bibitem{Tylka:2016aa}
\bibinfo{author}{J.~G. Tylka} and \bibinfo{author}{E.~Choueiri},
  \enquote{\bibinfo{title}{Soundfield navigation using an array of higher-order
  ambisonics microphones}}, in \emph{\bibinfo{booktitle}{Audio Engineering
  Society Conference: 2016 AES International Conference on Audio for Virtual
  and Augmented Reality}}, \bibinfo{organization}{Audio Engineering Society}
  (\bibinfo{year}{2016}).

\bibitem{Berkhout:1993aa}
\bibinfo{author}{A.~J. Berkhout}, \bibinfo{author}{D.~de~Vries}, and
  \bibinfo{author}{P.~Vogel}, \enquote{\bibinfo{title}{Acoustic control by wave
  field synthesis}},  \bibinfo{journal}{The Journal of the Acoustical Society
  of America} \textbf{93}(5), \bibinfo{pages}{2764--2778}
  (\bibinfo{year}{1993}).

\bibitem{Druyvesteyn:1997aa}
\bibinfo{author}{W.~Druyvesteyn} and \bibinfo{author}{J.~Garas},
  \enquote{\bibinfo{title}{Personal sound}},  \bibinfo{journal}{Journal of the
  Audio Engineering Society} \textbf{45}(9), \bibinfo{pages}{685--701}
  (\bibinfo{year}{1997}).

\bibitem{Ward:2001aa}
\bibinfo{author}{D.~B. Ward} and \bibinfo{author}{T.~D. Abhayapala},
  \enquote{\bibinfo{title}{Reproduction of a plane-wave sound field using an
  array of loudspeakers}},  \bibinfo{journal}{IEEE Transactions on speech and
  audio processing} \textbf{9}(6), \bibinfo{pages}{697--707}
  (\bibinfo{year}{2001}).

\bibitem{Betlehem:2005aa}
\bibinfo{author}{T.~Betlehem} and \bibinfo{author}{T.~D. Abhayapala},
  \enquote{\bibinfo{title}{Theory and design of sound field reproduction in
  reverberant rooms}},  \bibinfo{journal}{The Journal of the Acoustical Society
  of America} \textbf{117}(4), \bibinfo{pages}{2100--2111}
  (\bibinfo{year}{2005}).

\bibitem{Radlovic:2000aa}
\bibinfo{author}{B.~D. Radlovic}, \bibinfo{author}{R.~C. Williamson}, and
  \bibinfo{author}{R.~A. Kennedy}, \enquote{\bibinfo{title}{Equalization in an
  acoustic reverberant environment: Robustness results}},
  \bibinfo{journal}{IEEE Transactions on Speech and Audio Processing}
  \textbf{8}(3), \bibinfo{pages}{311--319} (\bibinfo{year}{2000}).

\bibitem{Williams:1999ab}
\bibinfo{author}{E.~G. Williams}, \emph{\bibinfo{title}{Fourier acoustics:
  sound radiation and nearfield acoustical holography}}
  (\bibinfo{publisher}{Elsevier}, \bibinfo{year}{1999}).

\bibitem{Torras-Rosell:2012aa}
\bibinfo{author}{A.~Torras-Rosell}, \bibinfo{author}{S.~Barrera-Figueroa}, and
  \bibinfo{author}{F.~Jacobsen}, \enquote{\bibinfo{title}{Sound field
  reconstruction using acousto-optic tomography}},  \bibinfo{journal}{The
  Journal of the Acoustical Society of America} \textbf{131}(5),
  \bibinfo{pages}{3786--3793} (\bibinfo{year}{2012}).

\bibitem{fernandez2013holographic}
\bibinfo{author}{E.~Fernandez-Grande}, \bibinfo{author}{A.~Torras-Rosell}, and
  \bibinfo{author}{F.~Jacobsen}, \enquote{\bibinfo{title}{Holographic
  reconstruction of sound fields based on the acousto-optic effect}}, in
  \emph{\bibinfo{booktitle}{INTER-NOISE and NOISE-CON Congress and Conference
  Proceedings}}, \bibinfo{organization}{Institute of Noise Control Engineering}
  (\bibinfo{year}{2013}), Vol. \bibinfo{volume}{247}, pp.
  \bibinfo{pages}{3181--3190}.

\bibitem{Ajdler:2006aa}
\bibinfo{author}{T.~Ajdler}, \bibinfo{author}{L.~Sbaiz}, and
  \bibinfo{author}{M.~Vetterli}, \enquote{\bibinfo{title}{The plenacoustic
  function and its sampling}},  \bibinfo{journal}{IEEE transactions on Signal
  Processing} \textbf{54}(10), \bibinfo{pages}{3790--3804}
  (\bibinfo{year}{2006}).

\bibitem{Haneda:1999aa}
\bibinfo{author}{Y.~Haneda}, \bibinfo{author}{Y.~Kaneda}, and
  \bibinfo{author}{N.~Kitawaki},
  \enquote{\bibinfo{title}{Common-acoustical-pole and residue model and its
  application to spatial interpolation and extrapolation of a room transfer
  function}},  \bibinfo{journal}{IEEE Transactions on Speech and Audio
  Processing} \textbf{7}(6), \bibinfo{pages}{709--717} (\bibinfo{year}{1999}).

\bibitem{Moiola:2011aa}
\bibinfo{author}{A.~Moiola}, \bibinfo{author}{R.~Hiptmair}, and
  \bibinfo{author}{I.~Perugia}, \enquote{\bibinfo{title}{Plane wave
  approximation of homogeneous helmholtz solutions}},
  \bibinfo{journal}{Zeitschrift f{\"u}r angewandte Mathematik und Physik}
  \textbf{62}(5), \bibinfo{pages}{809} (\bibinfo{year}{2011}).

\bibitem{Mignot:2013ab}
\bibinfo{author}{R.~Mignot}, \bibinfo{author}{G.~Chardon}, and
  \bibinfo{author}{L.~Daudet}, \enquote{\bibinfo{title}{Low frequency
  interpolation of room impulse responses using compressed sensing}},
  \bibinfo{journal}{IEEE/ACM Transactions on Audio, Speech, and Language
  Processing} \textbf{22}(1), \bibinfo{pages}{205--216} (\bibinfo{year}{2013}).

\bibitem{Antonello:2017aa}
\bibinfo{author}{N.~Antonello}, \bibinfo{author}{E.~De~Sena},
  \bibinfo{author}{M.~Moonen}, \bibinfo{author}{P.~A. Naylor}, and
  \bibinfo{author}{T.~van Waterschoot}, \enquote{\bibinfo{title}{Room impulse
  response interpolation using a sparse spatio-temporal representation of the
  sound field}},  \bibinfo{journal}{IEEE/ACM Transactions on Audio, Speech, and
  Language Processing} \textbf{25}(10), \bibinfo{pages}{1929--1941}
  (\bibinfo{year}{2017}).

\bibitem{Verburg:2018aa}
\bibinfo{author}{S.~A. Verburg} and \bibinfo{author}{E.~Fernandez-Grande},
  \enquote{\bibinfo{title}{Reconstruction of the sound field in a room using
  compressive sensing}},  \bibinfo{journal}{The Journal of the Acoustical
  Society of America} \textbf{143}(6), \bibinfo{pages}{3770--3779}
  (\bibinfo{year}{2018}).

\bibitem{Mignot:2013aa}
\bibinfo{author}{R.~Mignot}, \bibinfo{author}{L.~Daudet}, and
  \bibinfo{author}{F.~Ollivier}, \enquote{\bibinfo{title}{Room reverberation
  reconstruction: Interpolation of the early part using compressed sensing}},
  \bibinfo{journal}{IEEE Transactions on Audio, Speech, and Language
  Processing} \textbf{21}(11), \bibinfo{pages}{2301--2312}
  (\bibinfo{year}{2013}).

\bibitem{grande2019sound}
\bibinfo{author}{E.~F. Grande}, \enquote{\bibinfo{title}{Sound field
  reconstruction in a room from spatially distributed measurements}}, in
  \emph{\bibinfo{booktitle}{23rd International Congress on Acoustics}},
  \bibinfo{organization}{German Acoustical Society (DEGA)}
  (\bibinfo{year}{2019}), pp. \bibinfo{pages}{4961--68}.

\bibitem{Kim:2007aa}
\bibinfo{author}{S.-J. Kim}, \bibinfo{author}{K.~Koh},
  \bibinfo{author}{M.~Lustig}, \bibinfo{author}{S.~Boyd}, and
  \bibinfo{author}{D.~Gorinevsky}, \enquote{\bibinfo{title}{An interior-point
  method for large-scale $\ell_1 $-regularized least squares}},
  \bibinfo{journal}{IEEE journal of selected topics in signal processing}
  \textbf{1}(4), \bibinfo{pages}{606--617} (\bibinfo{year}{2007}).

\bibitem{bertalmio2000image}
\bibinfo{author}{M.~Bertalmio}, \bibinfo{author}{G.~Sapiro},
  \bibinfo{author}{V.~Caselles}, and \bibinfo{author}{C.~Ballester},
  \enquote{\bibinfo{title}{Image inpainting}}, in
  \emph{\bibinfo{booktitle}{Proceedings of the 27th annual conference on
  Computer graphics and interactive techniques}}, \bibinfo{organization}{ACM
  Press/Addison-Wesley Publishing Co.} (\bibinfo{year}{2000}), pp.
  \bibinfo{pages}{417--424}.

\bibitem{Liu:2018aa}
\bibinfo{author}{G.~Liu}, \bibinfo{author}{F.~A. Reda}, \bibinfo{author}{K.~J.
  Shih}, \bibinfo{author}{T.-C. Wang}, \bibinfo{author}{A.~Tao}, and
  \bibinfo{author}{B.~Catanzaro}, \enquote{\bibinfo{title}{Image inpainting for
  irregular holes using partial convolutions}}, in
  \emph{\bibinfo{booktitle}{Proceedings of the European Conference on Computer
  Vision (ECCV)}} (\bibinfo{year}{2018}), pp. \bibinfo{pages}{85--100}.

\bibitem{freeman2002example}
\bibinfo{author}{W.~T. Freeman}, \bibinfo{author}{T.~R. Jones}, and
  \bibinfo{author}{E.~C. Pasztor}, \enquote{\bibinfo{title}{Example-based
  super-resolution}},  \bibinfo{journal}{IEEE Computer graphics and
  Applications} \textbf{22}(2), \bibinfo{pages}{56--65} (\bibinfo{year}{2002}).

\bibitem{zhang2018residual}
\bibinfo{author}{Y.~Zhang}, \bibinfo{author}{Y.~Tian},
  \bibinfo{author}{Y.~Kong}, \bibinfo{author}{B.~Zhong}, and
  \bibinfo{author}{Y.~Fu}, \enquote{\bibinfo{title}{Residual dense network for
  image super-resolution}}, in \emph{\bibinfo{booktitle}{Proceedings of the
  IEEE Conference on Computer Vision and Pattern Recognition}}
  (\bibinfo{year}{2018}), pp. \bibinfo{pages}{2472--2481}.

\bibitem{Ronneberger:2015aa}
\bibinfo{author}{O.~Ronneberger}, \bibinfo{author}{P.~Fischer}, and
  \bibinfo{author}{T.~Brox}, \enquote{\bibinfo{title}{U-net: Convolutional
  networks for biomedical image segmentation}}, in
  \emph{\bibinfo{booktitle}{International Conference on Medical image computing
  and computer-assisted intervention}}, \bibinfo{organization}{Springer}
  (\bibinfo{year}{2015}), pp. \bibinfo{pages}{234--241}.

\bibitem{he2016deep}
\bibinfo{author}{K.~He}, \bibinfo{author}{X.~Zhang}, \bibinfo{author}{S.~Ren},
  and \bibinfo{author}{J.~Sun}, \enquote{\bibinfo{title}{Deep residual learning
  for image recognition}}, in \emph{\bibinfo{booktitle}{Proceedings of the IEEE
  conference on computer vision and pattern recognition}}
  (\bibinfo{year}{2016}), pp. \bibinfo{pages}{770--778}.

\bibitem{chollet2017xception}
\bibinfo{author}{F.~Chollet}, \enquote{\bibinfo{title}{Xception: Deep learning
  with depthwise separable convolutions}}, in
  \emph{\bibinfo{booktitle}{Proceedings of the IEEE conference on computer
  vision and pattern recognition}} (\bibinfo{year}{2017}), pp.
  \bibinfo{pages}{1251--1258}.

\bibitem{nassif2019speech}
\bibinfo{author}{A.~B. Nassif}, \bibinfo{author}{I.~Shahin},
  \bibinfo{author}{I.~Attili}, \bibinfo{author}{M.~Azzeh}, and
  \bibinfo{author}{K.~Shaalan}, \enquote{\bibinfo{title}{Speech recognition
  using deep neural networks: A systematic review}},  \bibinfo{journal}{IEEE
  Access} \textbf{7}, \bibinfo{pages}{19143--19165} (\bibinfo{year}{2019}).

\bibitem{wang2017tacotron}
\bibinfo{author}{Y.~Wang}, \bibinfo{author}{R.~Skerry-Ryan},
  \bibinfo{author}{D.~Stanton}, \bibinfo{author}{Y.~Wu}, \bibinfo{author}{R.~J.
  Weiss}, \bibinfo{author}{N.~Jaitly}, \bibinfo{author}{Z.~Yang},
  \bibinfo{author}{Y.~Xiao}, \bibinfo{author}{Z.~Chen},
  \bibinfo{author}{S.~Bengio}, \emph{et~al.},
  \enquote{\bibinfo{title}{Tacotron: Towards end-to-end speech synthesis}},
  \bibinfo{journal}{Proc. Interspeech 2017} \bibinfo{pages}{4006--4010}
  (\bibinfo{year}{2017}).

\bibitem{international2015recommendation}
\bibinfo{author}{I.~T. Union}, \enxquote{\bibinfo{title}{Recommendation itu-r
  bs. 1116-3: Methods for the subjective assessment of small impairments in
  audio systems}}  (\bibinfo{year}{2015}).

\bibitem{Note1}
\bibinfo{note}{See supplementary material at \protect \url
  {github.com/francesclluis/sound-field-neural-network}}.

\bibitem{jacobsen2013fundamentals}
\bibinfo{author}{F.~Jacobsen} and \bibinfo{author}{P.~M. Juhl},
  \emph{\bibinfo{title}{Fundamentals of general linear acoustics}}
  (\bibinfo{publisher}{John Wiley \& Sons}, \bibinfo{year}{2013}).

\bibitem{drozdzal2016importance}
\bibinfo{author}{M.~Drozdzal}, \bibinfo{author}{E.~Vorontsov},
  \bibinfo{author}{G.~Chartrand}, \bibinfo{author}{S.~Kadoury}, and
  \bibinfo{author}{C.~Pal}, \enquote{\bibinfo{title}{The importance of skip
  connections in biomedical image segmentation}}, in
  \emph{\bibinfo{booktitle}{Deep Learning and Data Labeling for Medical
  Applications}}  (\bibinfo{publisher}{Springer}, \bibinfo{year}{2016}), pp.
  \bibinfo{pages}{179--187}.

\bibitem{isola2017image}
\bibinfo{author}{P.~Isola}, \bibinfo{author}{J.-Y. Zhu},
  \bibinfo{author}{T.~Zhou}, and \bibinfo{author}{A.~A. Efros},
  \enquote{\bibinfo{title}{Image-to-image translation with conditional
  adversarial networks}}, in \emph{\bibinfo{booktitle}{Proceedings of the IEEE
  conference on computer vision and pattern recognition}}
  (\bibinfo{year}{2017}), pp. \bibinfo{pages}{1125--1134}.

\bibitem{wang2004image}
\bibinfo{author}{Z.~Wang}, \bibinfo{author}{A.~C. Bovik},
  \bibinfo{author}{H.~R. Sheikh}, \bibinfo{author}{E.~P. Simoncelli},
  \emph{et~al.}, \enquote{\bibinfo{title}{Image quality assessment: from error
  visibility to structural similarity}},  \bibinfo{journal}{IEEE transactions
  on image processing} \textbf{13}(4), \bibinfo{pages}{600--612}
  (\bibinfo{year}{2004}).

\bibitem{Landau:1967aa}
\bibinfo{author}{H.~J. Landau}, \enquote{\bibinfo{title}{Necessary density
  conditions for sampling and interpolation of certain entire functions}},
  \bibinfo{journal}{Acta Mathematica} \textbf{117}(1), \bibinfo{pages}{37--52}
  (\bibinfo{year}{1967}).

\bibitem{farina2000simultaneous}
\bibinfo{author}{A.~Farina}, \enquote{\bibinfo{title}{Simultaneous measurement
  of impulse response and distortion with a swept-sine technique}}, in
  \emph{\bibinfo{booktitle}{Audio Engineering Society Convention 108}},
  \bibinfo{organization}{Audio Engineering Society} (\bibinfo{year}{2000}).

\bibitem{iso2008acoustics}
\bibinfo{author}{I.~3382-2}, \enxquote{\bibinfo{title}{Acoustics —
  measurement of room acoustic parameters — part 2: Reverberation time in
  ordinary rooms}}  (\bibinfo{year}{2008}).

\bibitem{Ulyanov:2018aa}
\bibinfo{author}{D.~Ulyanov}, \bibinfo{author}{A.~Vedaldi}, and
  \bibinfo{author}{V.~Lempitsky}, \enquote{\bibinfo{title}{Deep image prior}},
  in \emph{\bibinfo{booktitle}{Proceedings of the IEEE Conference on Computer
  Vision and Pattern Recognition}} (\bibinfo{year}{2018}), pp.
  \bibinfo{pages}{9446--9454}.

\bibitem{Goodfellow-et-al-2016}
\bibinfo{author}{I.~Goodfellow}, \bibinfo{author}{Y.~Bengio}, and
  \bibinfo{author}{A.~Courville}, \emph{\bibinfo{title}{Deep Learning}}
  (\bibinfo{publisher}{MIT Press}, \bibinfo{year}{2016})
  \bibinfo{note}{\url{http://www.deeplearningbook.org}}.

\end{thebibliography}




\end{document}